\documentclass[twocolumn,prb,preprintnumbers,amsmath,amssymb,superscriptaddress]{revtex4}

\usepackage{color}
\definecolor{darkgreen}{rgb}{0,0.5,0}
\definecolor{blue}{rgb}{0,0,0.8}
\definecolor{lightblue}{rgb}{0.93,0.96,1}
\definecolor{darkblue}{rgb}{0.,0.,0.6}
\usepackage[colorlinks,linkcolor=blue,citecolor=blue,urlcolor=blue]{hyperref}

\usepackage{bm}% bold math
\usepackage{chemarr,color}
\usepackage{ifthen}  % Conditional
\usepackage{graphicx,subfigure}
\usepackage{placeins}

\renewcommand\vec[1]{\boldsymbol{\ensuremath{#1}}}
\def\lim{\mathop{\rm lim}}

\def\dd{\text{d}}

\begin{document}

\title{Stochastic theory of large-scale enzyme-reaction networks: finite copy number corrections to rate equation models\footnote{ Paper published in: \\[2mm]  \href{http://link.aip.org/link/?JCP/133/195101}{J. Chem. Phys. {\bf 133}, 195101 (2010)} \\ DOI: \href{http://dx.doi.org/10.1063/1.3505552}{10.1063/1.3505552} } }

\author{Philipp Thomas}
\affiliation{Department of Physics, Humboldt University of Berlin, Newtonstr. 15, D-12489 Berlin, Germany}

\author{Arthur V. Straube}
\affiliation{Department of Physics, Humboldt University of Berlin, Newtonstr. 15, D-12489 Berlin, Germany}

\author{Ramon Grima}
\affiliation{School of Biological Sciences, University of Edinburgh, Edinburgh EH9 3JR, United Kingdom}

\begin{abstract}
Chemical reactions inside cells occur in compartment volumes in the range of atto- to femtolitres. Physiological concentrations realized in such small volumes imply low copy numbers of interacting molecules with the consequence of considerable fluctuations in the concentrations. In contrast, rate equation models are based on the implicit assumption of infinitely large numbers of interacting molecules, or equivalently, that reactions occur in infinite volumes at constant macroscopic concentrations. In this article we compute the finite-volume corrections (or equivalently the finite copy number corrections) to the solutions of the rate equations for chemical reaction networks composed of arbitrarily large numbers of enzyme-catalyzed reactions which are confined inside a small sub-cellular compartment. This is achieved by applying a mesoscopic version of the quasi-steady state assumption to the exact Fokker-Planck equation associated with the Poisson Representation of the chemical master equation. The procedure yields impressively simple and compact expressions for the finite-volume corrections. We prove that the predictions of the rate equations will always underestimate the actual steady-state substrate concentrations for an enzyme-reaction network confined in a small volume. In particular we show that the finite-volume corrections increase with decreasing sub-cellular volume, decreasing Michaelis-Menten constants and increasing enzyme saturation. The magnitude of the corrections depends sensitively on the topology of the network. The predictions of the theory are shown to be in excellent agreement with stochastic simulations for two types of networks typically associated with protein methylation and metabolism.
\end{abstract}

\maketitle

\section{Introduction}

Recent years have seen a distinctive surge in the formulation and application of stochastic models of biochemical reaction kinetics. This trend has resulted from a deeper, ongoing appreciation of the conditions characteristic of the intracellular environment \cite{GrimaSchnell} and of their dissimilarity from \emph{in vitro} conditions. Typical \emph{in vivo} concentrations are in the range of nanomolar to millimolar; such concentrations realized in a macroscopic volume imply very large copy numbers of interacting molecules whereas the same concentrations in the small volume of a cell frequently implies copy numbers ranging from few tens to at most few thousands (for a detailed experimental protein abundance study see, for example, Ref.~\onlinecite{Ishihama2008}). Reaction kinetics is inherently a stochastic process;\cite{Gillespierev} this noisiness is not apparent in macroscopic conditions due to an implicit averaging over a very large number of molecules but cannot be overlooked when we are studying the kinetics of a system in which the copy number of at least one species is small. This is frequently the case of intracellular kinetics.

The introduction of the stochastic simulation algorithm by Gillespie \cite{Gillespie1} has popularized the numerical study of stochastic reaction kinetics. However to-date the analytical study of the properties of such systems has received comparatively very little attention principally because the mathematical formalism of stochastic kinetics [i.e., chemical master equations, (CMEs)] is very different than that of deterministic kinetics [i.e., rate equations (REs) which are based on ordinary differential equations] and is less amenable to analysis. This problem is augmented by the fact that many biological networks of interest are considerably large.

One of the main analytical methods for systematically exploring the stochastic properties of these networks has been the linear-noise approximation.\cite{VanKampen, ElfEhrenberg} The advantage of this method is the relative ease with which one can compute the magnitude of intrinsic noise (i.e., coefficients of variation and Fano factors). The major drawback is that the linear-noise approximation gives only meaningful results provided the copy number of molecules is not small or to be more precise it is correct in the limit of infinitely large reaction volumes, i.e., the same limit in which the REs are valid. Since intracellular reactions occur in the opposite limit of small volumes, it is highly desirable to calculate the finite-volume corrections to the concentrations and moments of intrinsic noise.

Developing a theory of finite-volume corrections presents a considerable analytical challenge. One way to obtain the latter is via the system-size expansion of the CME.\cite{VanKampen} The linear-noise approximation comes about by evaluating the first term (of order $V^0$ where $V$ is the reaction volume) in this expansion. The next term is proportional to $V^{-1/2}$ and hence consideration of this term will necessarily give a finite-volume correction. Grima\cite{Grima1,Grima2} calculated the first such corrections for the mean concentrations of species involved in single-substrate enzyme reactions and recently also for a general chemical reaction network of arbitrary complexity.\cite{Grima3} The general result is that the kinetics of a reaction pathway confined in a volume $V$ can be described by the usual REs plus new terms which are proportional to $V^{-1}$. These equations are referred to as effective mesoscopic rate equations (EMREs). The differences between the solution of EMREs and the corresponding REs for finite volumes stems from a coupling between the mean concentrations and the fluctuations about them. EMREs can be explicitly solved for pathways characterized by a handful of chemical species, but otherwise one has to resort to numerical solution. In the latter cases, one obtains a solution at the expense of losing the insight which typically comes from analytical results.

In this article we develop an alternative, powerful method of calculating finite-volume corrections to the solutions of the REs. Note that in the context of this article, finite-volume corrections exclusively refer to corrections to the mean concentrations not to the moments of intrinsic noise. The method is based on the Poisson representation of the CME rather than the system-size expansion used in the derivation of EMREs. This method unlike the system-size expansion has been applied to study systems of biochemical or biological relevance in only a handful of cases (see, for example, Refs.~\onlinecite{Fleck} and \onlinecite{Drummond}) but as we shall show it is a tool with great potential for this field. We focus on chemical reaction networks which are composed of enzyme-catalyzed reactions, a commonly encountered case in intracellular biochemistry.\cite{Alberts} We show that when the timescales of complex and substrate fluctuations are well-separated, it is possible to obtain explicit and impressively simple equations for the finite-volume corrections. For simple reactions these corrections are shown to be the same as given by the EMRE. The distinct advantage of the new method over the EMRE is that it provides analytically simple results even for complex networks with hundreds or thousands of species. This is inherently possible because of the large reduction in the effective dimensionality of the CME when timescales are well separated.

The paper is organized as follows. In Section II, we derive the Poisson Representation for a general enzyme-reaction network and use the resulting Fokker-Planck equation (FPE) to obtain an exact Liouville equation encoding all information about deviations from the deterministic solution of the REs. In Section III we show that in the limit of well-separated timescales of complex and substrate species, the Liouville equation simplifies to a compact approximate form. This is used in Section IV to compute explicit expressions for the finite-volume corrections of a general network. In the latter section we show that the corrections for a simple Michaelis-Menten type reaction agree with those previously derived using the EMRE formalism. More importantly we apply the theoretical results to two common types of large-scale networks and confirm the predictions using simulations. We finish by a discussion in section V.

\section{The Poisson Representation for the enzyme reaction network}

In this section we use the Poisson Representation to derive a general FPE for an enzyme reaction network. The latter while being exactly equivalent to the CME is much more amenable to analysis and hence is a very convenient starting point for detailed calculation purposes.

We consider a generic type of enzyme network composed of two major types of chemical processes: (i) the input of a substrate species $A_0$ into a subcellular compartment; (ii) the transformation of $A_0$ into some final product $A_N$ via $N$ consecutive enzyme-catalyzed reactions of the type
\begin{equation}
 \label{eqn:singlestep}
 {E}_i+A_i\xrightleftharpoons[k^i_{-1}]{k_1} C_i \overset{k^i_2}{\longrightarrow} {E}_i+A_{i+1},
\end{equation}
where $A_i$, $C_i$ and $E_i$ denote the $i^{th}$ substrate, complex and enzyme species, respectively; the index $i$ takes values from 0 to $N-1$ and the $k's$ denote the relevant macroscopic rate constants. Note that we have assumed here that the bimolecular reaction rate, $k_1$, is the same for all substrates.  Note also that in such types of networks there are $N$ distinct substrate species, an equal number of distinct complex species and a number of enzyme species varying between 1 and $N$.  The freedom in choosing the number of enzyme species comes from the fact that an enzyme can generally bind to more than one type of substrate.

If we assume that we have well-mixed conditions inside the compartment then the instantaneous description of the state of a chemical system at time $t$ is simply given by the vector of the absolute number of molecules of each species, $\vec{n}=(\{n_{A_i}\},\{n_{E_i}\},\{n_{C_i}\})$. Since the mesoscopic kinetics are stochastic, a full description of the system is necessarily probabilistic and is achieved by defining the probability density function, $P=P(\vec{n},t)$ and its time-evolution equation, which is commonly referred to as the CME (Refs.~\onlinecite{VanKampen} and \onlinecite{Gardiner2})
\begin{align}
 \label{eqn:master_general}
 \partial_t P	=&  \sum_{i=0}^{N-1} \Pi_i P + {\Xi}  P,\\
 \Pi_i 		=& \, \frac{k_1}{V}(\Theta_{A_i} \Theta_{C_i}^{-1}-1){n_{A_i} n_{E_i}} \notag\\
		& + k_{-1}^i (\Theta_{C_i}\Theta_{A_i}^{-1}-1)n_{C_i} \notag \\
        & + k_2^i (\Theta_{C_i}\Theta_{A_{i+1}}^{-1}-1) n_{C_i},\\
 {\Xi} 	=& \;k_\text{in}V(\Theta_{A_0}^{-1}-1),
\end{align}
where $V$ is the compartment volume, {$\Xi$} is the contribution due to the input of substrate species $A_0$ into the system at a rate $k_{in}$ while $\Pi_i$ describes the $i^{th}$ catalytic reaction step, Eq. (\ref{eqn:singlestep}), in the reaction network. The CME is compactly expressed using van Kampen's step operators defined as $\Theta_{X_i}^{\pm 1} g(n_{X_i}) = g(n_{X_i} \pm 1)$.

Substantially, the CME depends only on the set of variables $\{n_{A_i}\}$ and $\{n_{C_i}\}$ since an enzyme molecule can be either in the free state or in the complexed state and hence the variables $\{n_{E_i}\}$ are redundant. We can express this conservation law by writing $n_{E_i}=E_T^i-\sum_j G_{ij} n_{C_j}$ where the matrix $G_{ij}$ is defined by construction to be
\begin{align}
\label{eqn:Gdefn}
 G_{ij}=(\vec{\hat{e}_i})_j=\left\{
\begin{array}{ll}
1, & \text{ if enzyme $i$ binds substrate $j$}\\
0, & \text{ otherwise}.
\end{array}\right.
\end{align}
The $N$-dimensional vector, $\vec{\hat{e}_i}$, is associated with the enzyme binding substrate in the $i$-th catalytic reaction. Its $j$-th entry is chosen to be equal to one if the enzyme can form a complex with the $j$-th substrate and zero otherwise. Hence the connectivity of the network is explicitly encoded in the form of these vectors.

We define the moment generating function, parameterized by the vector of continuous variables, $\vec{z}=(\{z_{A_i}\},\{z_{C_i}\})$, as
\begin{align}
  G(\vec{z})=\sum_{\vec{n}} \prod_{i} z_{A_i}^{n_{A_i}} z_{C_i}^{n_{C_i}} P(\vec{n}).
\end{align}
Multiplying the CME, Eq.~(\ref{eqn:master_general}), by $\prod_{i} z_{A_i}^{n_{A_i}} z_{C_i}^{n_{C_i}}$, performing the summation over all values of the variables $\vec{n}$ and expressing the resulting equation in terms of the moment generating function $G(\vec{z})$, we obtain the moment generating function equation
\begin{align}
 \label{eqn:mgf_eq}
 \partial_t G(\vec{z}) 	&= \left(k_1{\sum}_i\mathcal{R}^G_i + \mathcal{S}^G\right) G, \\
 \mathcal{S}^G 		&= k_\text{in} V (z_{A_0}-1), \\
 \label{eqn:RG}
 \mathcal{R}^G_i 		&= (z_{C_i}- z_{A_i}) \left( \hat{\mathcal{E}}_i \partial_z^{A_i} - K_M^i \partial_z^{A_i} \right) \notag \\
 & + K_2^i(z_{A_{i+1}}-z_{A_{i}}) \,\partial_z^{C_i},
\end{align}
where $\hat{\mathcal{E}}_i$ abbreviates $[E_T^i]-V^{-1}\sum z_{C_i} \partial_C^i$ and $[E_T^i]=E_T^i/V$ is the total enzyme concentration associated with the enzyme binding substrate in the $i^{th}$ catalytic reaction step. Furthermore we set $K_1^i=k_{-1}^i/k_1$, $K_2^i=k_{2}^i/k_1$, $K_M^i=K_1^i+K_2^i$ (commonly referred to as the Michaelis-Menten constant) and $\partial_z^X = \partial/\partial z_X$. Note that in Eq.~(\ref{eqn:RG}) all derivatives are to the right.

We proceed by making use of Gardiner's Poisson Representation. At the heart of this method is the assumption that the probability density function $P(\textbf{n},t)$ can be expanded as a superposition of multivariate uncorrelated Poissons \cite{Gardiner2,Gardiner1}
\begin{align}
\label{eqn:superpos}
P(\vec{n},t)=&\int \dd {\vec\alpha }\, \prod_i \frac{e^{-\alpha_{A_i}V}
(\alpha_{A_i}V)^{n_{A_i}}}{n_{A_i}!} \nonumber \\
&\quad\qquad\times \frac{e^{-\alpha_{C_i}V} (\alpha_{C_i}V)^{n_{C_i}}}{n_{C_i}!} f(\vec\alpha,t),
\end{align}
where the function $f(\vec\alpha,t)$ is usually referred to as the quasi-probability density function. The vector $\vec \alpha$ is defined to be $(\{\alpha_{A_i}\},\{\alpha_{C_i}\})$. It has been shown that this superposition always exists if the range of $\vec\alpha$ is extended to the complex plane by analytic continuation of the Poisson kernel. We have explicitly introduced $V$ in our superposition definition, which is not customarily done in the Poisson Representation. In doing so, we obtain the representation in intensive variables, i.e. in units of concentrations, as those encountered in the theory of rate equations. The above expansion is equivalent to writing the moment generating function as
\begin{align}
\label{eqn:superpos_G}
G(\vec{z},t)=\int \dd{\vec\alpha } \, f(\vec\alpha,t) e^{\sum (z_{A_i}-1) \alpha_{A_i} V} e^{\sum (z_{C_i}-1)\alpha_{C_i}V}
\end{align}
It follows from Eqs.~(\ref{eqn:mgf_eq}) and (\ref{eqn:superpos_G}) (see Appendix \ref{app:rp} for details)
\begin{align}
 \label{eqn:fFPE}
 -\partial_t f	=& \; (\mathcal{R}+\mathcal{S}) f, \\
 \mathcal{S} 	=& \; k_\text{in} \partial_{A}^0,\\
 \mathcal{R}   	=&  \, k_1 \sum^{N-1}_{i=0} \mathcal{R}_i,\notag\\
 \mathcal{R}_i 	=& \; (\partial_{C}^i-\partial_{A}^i)\left(\alpha_{A_i} \mathcal{E}_i -K_M^i \alpha_{C_i} \right) \notag \\
  & +K_2^i (\partial_{A}^{i+1}-\partial_{A}^{i}) \alpha_{C_i},
\end{align}
where we have utilized the notation $\nu=V^{-1/2}$, $\partial_{X}^i = \partial/\partial \alpha_{X_i}$. Note that $\alpha_{A_N}=\alpha_P$ is the variable for the product formed after $N$ catalytic steps. Note also that $\mathcal{E}_i$ is not a constant, but is given by the operator
\begin{align}
 \label{eqn:Eoperator}
 \mathcal{E}_i=[E_T^i]- {\sum}_{j} G_{ij} (1-\nu^2 \partial_{C_j})\alpha_{C_j},
\end{align}
which is essentially the Poisson representation of the conservation of total enzyme molecules. The above equation generally differs from the corresponding deterministic conservation law in terms of concentrations; the latter is described only by its average, while the former exhibits finite volume corrections due to the finite copy number of enzyme molecules. Given Eq.~(\ref{eqn:Eoperator}), we see that the Poisson representation, Eq.~(\ref{eqn:fFPE}), yields a Fokker-Planck equation in terms of substrate and complex variables. Note also that for the case of $N=1$ we obtain the representation for the single-substrate single-enzyme reaction, which is usually referred to as the Michaelis-Menten reaction.

This completes the derivation of the Poisson representation of our general enzyme-reaction network. Note that this is not the same FPE as that which arises from the system-size expansion method of van Kampen.\cite{VanKampen} In the latter case, the FPE is an approximation to the CME in the limit of large volumes whereas the FPE obtained from the Poisson Representation is exactly equivalent to the CME.

A further boon of the Poisson representation is that once we have calculated the moments of the continuous variables, $\alpha_{X_i}$, using the FPE, we can very easily find the corresponding moments of the copy number of molecules, $n_{X_i}$, using the following simple relationships:\cite{Gardiner1}
\begin{align}
\label{eqn:means}
 \frac{\langle n_{X_i} \rangle}{V} 	&= \langle \alpha_{X_i} \rangle, \\
 {\frac{\langle n_{X_i} n_{X_j} \rangle}{V^2}}	&= \langle \alpha_{X_i} \alpha_{X_j} \rangle + \frac{1}{V}\delta_{i,j} \langle \alpha_{X_i} \rangle,
\end{align}
where the angled brackets imply the statistical average: on the left hand side of the above equations these are given by $\langle .. \rangle = \int \text{d} \vec n \ .. \ P(\vec n,t)$ while those on the right hand side imply $\langle .. \rangle$ = $\int \text{d} \vec\alpha  \ ..  \ f(\vec\alpha,t)$.

\subsection{The mesoscopic equation}

The kinetics becomes deterministic in the macroscopic limit of infinitely large volumes. This can be easily verified by noting that all second-order derivatives in the FPE are multiplied by a factor proportional to the inverse of the volume. To compute the finite-volume corrections we will need to separate the mesoscopic and macroscopic evolution equations. We now show that this can be done by applying a suitable change of variables to the Poisson Representation. The macroscopic corresponds to a shot noise contribution, which agrees on average with the mean-field result i.e. with the solution of the REs for the substrate-enzyme network. The mesoscopic contribution reflects the non-equilibrium properties of the network due to the bimolecular character of the substrate-enzyme interaction, and depends parametrically on the mean-field expectation.

We express the deviation from the deterministic path by the following change of variables:
\begin{align}
 \label{eqn:ansatz}
 (\alpha_{A_i},\alpha_{C_i}) \rightarrow ([A_i](t) + \nu \epsilon_{A_i}, [C_i](t) + \nu \epsilon_{C_i})
\end{align}
We shall refer to $\epsilon_{X_i}$ as the mesoscopic correction to the deterministic, macroscopic concentration $[X_i](t)$ of species $X_i$. Note that the above equation has the same apparent form as the van Kampen (VK) ansatz,\cite{VanKampen,ElfEhrenberg} at the heart of the system-size expansion, but the context of the application is completely different. The VK ansatz is applied to the integer number of particles in the CME leading to an infinite series in powers of the inverse square root of the volume, which has to be truncated; the first term of this expansion (the one proportional to $V^0$) gives a linear FPE which is an approximation to the CME. In our case the change of variables, Eq.~(\ref{eqn:ansatz}), is applied on the FPE arising from the Poisson representation which leads to a finite series and allows for exact analytical treatment; as we shall show now, this divides the exact FPE into a macroscopic term and a term which captures \emph{all} deviations from the macroscopic.

The transformation applied to the FPE transforms the time derivative into
\begin{align}
\label{eqn:ram}
\left. \frac{\partial}{\partial t}\right|_{\vec\alpha} f(\vec\alpha,t) = &
  \left. \frac{\partial}{\partial t}\right|_{\vec\epsilon} f([\vec{X}]+\nu\vec\epsilon,t) \notag \\
  & + \left.\frac{d \vec\epsilon}{dt}\right|_{\vec\alpha} \cdot \nabla_{\vec\epsilon} \,f([\vec{X}]+\nu\vec\epsilon,t),
\end{align}
where $\partial/\partial t|_x$ denotes taking the derivative with $x$ held constant.
It follows from Eq.~(\ref{eqn:ansatz}) that ${d \vec\epsilon}/{dt}|_{\vec\alpha}=-\nu^{-1} {d [\vec{X}]}/{dt}$.
Finally by expressing the right hand side of Eq.~(\ref{eqn:fFPE}) in terms of the new variables $\vec{\epsilon}$ and equating the result to Eq.~(\ref{eqn:ram}), we find that the FPE takes the form
\begin{align}
\label{eqn:newF}
- \frac{\partial }{\partial t} g (\vec\epsilon)
= & k_1 \mathcal{L} g (\vec\epsilon) + \nu^{-1} \bigg( k_1 \mathcal{R}_{\text{macro}} + \mathcal{S}_{\text{macro}} \frac{}{} \notag \\
 &  - \frac{\partial [\vec{A}]}{\partial t}\cdot \nabla_{ \epsilon_A} - \frac{\partial [\vec{C}]}{\partial t} \cdot \nabla_{\epsilon_C} \bigg) g (\vec\epsilon),
\end{align}

where the relevant operators are
\begin{align}
& \mathcal{S}_{\text{macro}} = k_\text{in}\frac{\partial }{\partial \epsilon_{A_0}}, \\
& \mathcal{R}_{\text{macro}}=\sum_i ([E_i] [A_i]-K_M^i [C_i])\frac{\partial }{\partial \epsilon_{C_i}} \nonumber \\
& \qquad\qquad + \sum_i(K_1^i [C_i]-[E_i] [A_i])\frac{\partial }{\partial \epsilon_{A_i}}, \\
\label{eqn:Lmes}
& \mathcal{L} = {\sum}_i \big\{(\partial_{C}^i-\partial_{A}^i)\big( [E_i]\epsilon_{A_i} + [A_i]\delta\mathcal{E}_i + \nu\epsilon_{A_i}\delta\mathcal{E}_i  \nonumber \\
& \qquad\qquad -K_M^i \epsilon_{C_i} \big)+ K_2^i (\partial_{A}^{i+1}-\partial_{A}^{i}) \epsilon_{C_i} \big\}.
\end{align}
Note that the new probability density function necessarily satisfies $g(\vec\epsilon)\dd \vec\epsilon=f(\vec\alpha)\dd \vec\alpha$. Note also that the notation $\partial_{X}^i$ now denotes the derivative $\partial/\partial \epsilon_{X_i}$. The quantity $[E_i]=[E_T^i]-\sum_{j}G_{ij}[C_{j}]$ is the macroscopic concentration of the free enzyme species associated with binding substrate in the $i^{th}$ catalytic reaction step. The operator $\delta\mathcal{E}_i=-([E_i] - \mathcal{E}_i)/\nu=- \sum_j G_{ij} (\epsilon_{C_j}-\partial_C^j[C_j] -\nu \partial_C^j \epsilon_{C_j})$ is the contribution of the enzyme-operator Eq.~(\ref{eqn:Eoperator}).

Note that the resulting form of the FPE is clearly divided into two parts. In the macroscopic limit the terms proportional to $\nu^{-1}$ dominate and their sum must equate to zero - this leads to the macroscopic equations. It also then follows that the mesoscopic equation is simply given by
\begin{align}
\partial_\tau g(\vec\epsilon,t) &= - \mathcal{L} g(\vec\epsilon,t),
\end{align}
where we have renormalized time to $\tau=k_1 t$. Note also that due to the Poissonian nature of the substrate input process, it only contributes to the macroscopic part, a feature which is unique to the Poisson representation. We emphasize that up till this point, we have made no approximations and hence the resulting mesoscopic equation is exact.

\section{Adiabatic elimination of the complex species variables}
In this section we show how to rigorously eliminate the fast variables from our description. This will be done in two steps: on the macroscopic contribution of the FPE and on the mesoscopic contribution including terms up to order $\nu$ (i.e., finite volume corrections). As we shall see later on, the reduced mesoscopic description does depend on the reduced macroscopic description and hence the need to treat the latter first.

\subsection{The reduced macroscopic equations}
The macroscopic equations are obtained from Eq.~(\ref{eqn:newF}) by taking the limit $\nu \rightarrow 0$, leading to
\begin{align}
 \label{eqn:meanfield}
 \frac{ d [A_i]}{dt} &= k_{-1}^i [C_i]-k_1 [E_i] [A_i] + k_{2}^{i-1} [C_{i-1}]  + \delta_{i,0}k_\text{in}, \notag\\
 \frac{ d [C_i]}{dt} &= k_1 [E_i][A_i]-(k_2^i+k_{-1}^i) [C_i], \notag\\
 \frac{ d [A_N]}{dt} &= k_2^{N-1} [C_{N-1}], \notag\\
 [E_i] &= [E_T^i] - \sum_j G_{ij} [C_j].
\end{align}
These agree exactly with those that can be obtained from the RE approach. The elimination of the complex species from the macroscopic equations is a well-known procedure commonly referred to as the quasi-steady state approximation (QSSA).\cite{Segel,CornishBowden1995} Briefly speaking the approximation is tantamount to assuming that the complex equilibrates on a much shorter timescale than the substrate. This is implemented by imposing the approximation ${d[C_i]}/{dt}=0$ on the macroscopic equations. The resulting reduced equations (commonly referred to as the Briggs-Haldane equations) are then given by replacing $[C_i]$ in the full time-evolution equations for the substrate concentrations by
\begin{align}
 \label{eqn:MFss_cond}
 [C_i]=\frac{[E_i][A_i]}{K_M^i}.
\end{align}

\subsection{The reduced mesoscopic equation}
Now we are interested in deriving the reduced mesoscopic equation corresponding to the reduced macroscopic equations that we just considered. Time-scale separation on the mesoscopic scale is non-trivial because of the inherent correlations between the mesoscopic fluctuations of the various species. Our presentation shall be as follows. First we shall show that the mesoscopic Liouvillian, Eq.~(\ref{eqn:Lmes}), can be generally cast into an asymptotic form of the interaction representation which is typically encountered in the theory of adiabatic elimination of fast fluctuating variables.\cite{Gardiner3, Haake} The latter yields a particularly simple result for the reduced mesoscopic equation.

We will now show that the Liouvillian can be rewritten in the general form
\begin{align}
 \label{eqn:asymptoticform}
  \mathcal{L}(\gamma) = \gamma \mathcal{L}_1+\gamma^{1/2} \mathcal{L}_2 + \mathcal{L}_3,
\end{align}
where $\gamma^{-1}$ is the characteristic fast timescale of the complex fluctuations which will be specified later on. We now proceed to derive the operators $\mathcal{L}_1$, $\mathcal{L}_2$, $\mathcal{L}_3$ for the general enzyme-reaction network under study. We start by grouping all terms containing only the pair of complex variables $(\{\epsilon_{C_i}\}, \{\partial_C^i\})$ into $\mathcal{L}_1$, terms concerning solely the substrate ones $(\{\epsilon_{A_i}\}, \{\partial_A^i\})$ into $\mathcal{L}_3$, while treating the remaining terms as interaction $\mathcal{L}_2$. Thus we have
\begin{align}
 \mathcal{L}_1= &\sum_i   \mathcal{L}_1^{(i)}, \ \ \mathcal{L}_2=\sum_i   \mathcal{L}_2^{(i)}, \ \ \mathcal{L}_3=\sum_i   \mathcal{L}_3^{(i)}, \\
 \mathcal{L}_1^{(i)} = &\;\partial_C^i ( [A_i]\delta\mathcal{E}_i -K_M^i \epsilon_{C_i}), \label{eqn:candidate_L1}\\
 \mathcal{L}_2^{(i)} =&-
      \partial_A^i ( [A_i]\delta\mathcal{E}_i -K_M^i \epsilon_{C_i})
    + K_2^i (\partial_A^{i+1}-\partial_A^i)\epsilon_{C_i} \nonumber \\
    & + \nu(\partial_{C}^i-\partial_{A}^i) \epsilon_{A_i}\delta\mathcal{E}_i,  \\
 \mathcal{L}_3^{(i)} =& -\partial_A^i [E_i]\epsilon_{A_i}.
\end{align}
It is instructive to rescale all complex variables by their characteristic timescale $\gamma$
\begin{align}
 z_i = \gamma^{1/2} \epsilon_{C_i}, \ \ x_i=\epsilon_{A_i}.
\end{align}
The operator $\mathcal{L}_3$ is trivially obtained
\begin{align}
 \label{eqn:L3}
 \mathcal{L}_3^{(i)}
  =& -\partial_{x}^i [E_i] x_i.
\end{align}
Note that $\partial_{x}^i$ denotes the derivative $\partial/\partial x_i$ whereas $\partial_{X}^i$ stands for the derivative $\partial/\partial \epsilon_{X_i}$. Next we observe that the enzyme operator transforms as
\begin{align}
 \label{eqn:asymptotic_E}
 \gamma^{1/2}\delta\mathcal{E}_i=-\sum_j G_{ij} (z_j-\gamma\partial_z^j[C_j] -\nu \gamma^{1/2} \partial_z^j {z_j}).
\end{align}
Plugging this into Eq.~(\ref{eqn:candidate_L1}) and putting $\mathcal{L}_1^{(i)}\to \gamma\mathcal{L}_1^{(i)}$ we find that the last term in (\ref{eqn:asymptotic_E}) can be asymptotically neglected. Hence the dominant contribution in the limit of large $\gamma$ is given by
\begin{align}
 \label{eqn:L1}
 \mathcal{L}_1^{(i)}
  =& - \partial_{z}^i {\sum}_j M_{ij} z_j + \partial_{z}^i {\sum}_j D_{ij} \partial_{z}^j,
\end{align}
where we have utilized the abbreviations
\begin{align}
\begin{array}{c}
{J_{ij}}={([A_i]G_{ij}+K_M^i\delta_{ij})}, \ \  M_{ij}={\gamma}^{-1}J_{ij},  \\[2mm]
\ \ D_{ij}= G_{ij} {[A_i] [C_j]}.
\end{array}
\end{align}

Thus the asymptotic form of the complex fluctuations is described by an Ornstein-Uhlenbeck process, centered on the deterministic expectation Eq.~(\ref{eqn:MFss_cond}). It is clear, by virtue of the fluctuation-dissipation theorem, that the matrix $\underbar{J}$ (constant matrices, i.e., those independent of the $\epsilon_{A_i}$ and $\epsilon_{C_i}$ variables and of the associated partial derivatives, are underlined throughout the rest of the article) must correspond to the Jacobian of the complex species while the matrix $\underbar{D}$ determines the strength of the fluctuations in the Poisson representation. Note that throughout the article by Jacobian of a species we mean the negative of the Jacobian matrix as obtained from the macroscopic rate equations of that species with the time scaling $\tau = k_1 t$.

The characteristic timescale $\gamma^{-1}$ can be inferred from the explicit form of the Jacobian. We choose $\gamma\equiv\text{tr}(\underbar{J})$ giving the relaxation rate of the complex vector $\vec{z}$. Consequently $\underbar{M}$ is finite, as required by the stability of $\mathcal{L}_1$, and is given by the Jacobian of unit trace.
Equally we observe that the Jacobian of the substrate vector $\vec{x}$ is given by the diagonal matrix $\underbar{E}=\text{diag}([E_i])$, such that its relaxation rate is proportional to its trace. Thus under the condition where $\gamma\gg\text{tr}(\underbar{E})$, we can assume that the  timescales of substrate and complexes are well-separated. The result can be interpreted as follows: $\mathcal{L}_1$ represents the fast nonequilibrium fluctuations of the complex and $\mathcal{L}_3$ reflects the slow Poissonian decay of the substrate fluctuations.

Finally we turn our attention to the interaction, we put $\mathcal{L}_2\to\gamma^{1/2}\mathcal{L}_2$ and rearrange to emphasize its dependence on the complex variables
\begin{align}
 \label{eqn:L2_int}
 \mathcal{L}_2^{(i)}	&= {\sum}_j \chi_{ij} z_j + {\sum}_j \varphi_{ij} \partial_{z}^j, \notag\\
 \chi_{ij}		&=  \partial_{x}^i M_{ij} + \frac{1}{\gamma} \left( \delta_{ij} K_2^i (\partial_{x}^{i+1}-\partial_{x}^{i}) + \nu \partial_{x}^{i} x_i \right),\notag\\
 \varphi_{ij}		&=  \left(\delta_{ij} [E_i] x_i -\partial_{x}^{i} D_{ij} -{\nu}  \partial_{x}^{i} x_i G_{ij} [C_j] \right).
\end{align}
For notational convenience, we have considered only terms proportional to $\vec{z}$ and $\nabla_z$, namely those which give non-vanishing contributions in the following. We emphasize that Eqs.~(\ref{eqn:L3}), (\ref{eqn:L2_int}) together with (\ref{eqn:L1}) yield the contribution to the mesoscopic equation under the condition of well-separated substrate- and complex timescales.

It is clear that the reduced substrate density function $h(\vec{x})=\int \dd \vec{z} g(\vec{x},\vec{z})$, cannot simply be deduced from $\mathcal{L}_3$, but must retain a contribution from the inherent coupling to the rapid fluctuations of the complex species. However, it is known \cite{Gardiner2} that in the limit of well-separated timescales, a particularly simple result can be obtained by projecting the evolution equation on the steady-state of the fast variables. The technique is the adiabatic elimination of the fast variables and can be accomplished by the use of the projector $Pg=\pi(\vec{z})\int\dd \vec{z} g(\vec{x},\vec{z})$, with $\pi$ being the steady-state distribution of the fast variable vector $\vec{z}$, determined by the steady-state condition $\mathcal{L}_1\pi=0$.

The applicability of the method follows directly from the asymptotic form Eq.~(\ref{eqn:asymptoticform}), which has been derived here explicitly for the large scale enzyme-reaction networks under consideration, together with the conditions
$ P \mathcal{L}_3   = \mathcal{L}_3 P,
  P\mathcal{L}_1    = \mathcal{L}_1 P=0,
  P \mathcal{L}_2 P = 0$,\cite{Gardiner2}
which can be easily verified from Eqs.~(\ref{eqn:L3}), (\ref{eqn:L1}), and (\ref{eqn:L2_int}).
The result in the limit of large $\gamma$ has been formally derived by Gardiner \cite{Gardiner2} and reads
\begin{align}
  -\partial_t (Pg) = \lim_{\gamma\rightarrow\infty} P \mathcal{L}(\gamma)g= (\mathcal{L}_3 - P\mathcal{L}_2\mathcal{L}_1^{-1}\mathcal{L}_2) (Pg).
\end{align}
Integration over $\vec{z}$ yields the reduced evolution equation
\begin{align}
 \label{eqn:red_eqR}
 -\partial_t h(\vec{x},t)&=\mathcal{L}' h(\vec{x},t), \\
 \label{eqn:asymptoticLiouvillian}
 \mathcal{L}'&=  \lim_{\gamma\rightarrow\infty} \mathcal{L}(\gamma)= \mathcal{L}_3 - \langle \mathcal{L}_2\mathcal{L}_1^{-1}\mathcal{L}_2 \rangle_\pi.
\end{align}
The mesoscopic description is now reduced to that of the substrate only. The angled brackets $\langle \cdot \rangle_\pi$ denote the trace $\int \text{d} \vec{z} \cdot \pi(\vec{z})$ taken over the rapid steady-state fluctuations of the complex species. We now substitute Eq.~(\ref{eqn:L2_int}) into Eq.~(\ref{eqn:asymptoticLiouvillian})
\begin{align}
 \mathcal{L'}
      = \mathcal{L}_3
         - \sum_{ij} \left(\chi \langle \vec{z} \mathcal{L}_1^{-1} \vec{z}^T\rangle_\pi \chi^T \right)_{ij} \nonumber \\
         - \sum_{ij} \left(\chi \langle \vec{z} \mathcal{L}_1^{-1} \nabla_{z}^T\rangle_\pi \varphi^T\right)_{ij}.
 \label{eqn:fsolution0}
\end{align}
The explicit form of the correlators involved has been derived in Appendix \ref{app:transport_coeffs} using the steady-state condition. The result is
\begin{align}
 \label{eqn:transport_coeff}
\begin{array}{c}
 -\langle \vec{z}\mathcal{L}_1^{-1} \vec{z}^T\rangle_\pi = \underbar{S} \equiv \underbar{M}^{-1} \underbar{D}^T \underbar{M}^{-T},  \\[2mm]
-\langle \vec{z}\mathcal{L}_1^{-1} \nabla_{z} ^T \rangle_\pi = \underbar{M}^{-1}.
\end{array}
\end{align}

Hence using Eq.~(\ref{eqn:fsolution0}) together with Eq.~(\ref{eqn:transport_coeff}), we get the final form for the reduced Liouvillian in a convenient matrix form, which yields the mesoscopic implementation of the QSSA
\begin{align}
 \mathcal{L'} = \mathcal{L}_3 +  \sum_{ij}(\chi \underbar{S} \chi^T)_{ij} +\sum_{ij}(\chi \underbar{M}^{-1} \varphi^T)_{ij}.
 \label{eqn:fsolution}
\end{align}

\section{Finite volume corrections}

In this section we will use the reduced mesoscopic description to compute the finite volume corrections to the macroscopic steady-state concentrations of all substrate species in the network. By definition, the concentration of species $A_\alpha$ according to the stochastic model is given by $\rho_\alpha = \langle n_\alpha/V \rangle$. Using Eq.~(\ref{eqn:means}) together with Eq.~(\ref{eqn:ansatz}) it is straightforward to show that
\begin{equation}
\label{eqn:dens_pred_stoch}
\rho_\alpha = [A_\alpha] + \nu {\langle x_\alpha \rangle}.
\end{equation}
Hence the size of the finite-volume correction to the macroscopic concentration of species $A_\alpha$ is simply given by $\langle x_\alpha \rangle$. The time-evolution equation for the latter can be computed directly from our reduced mesoscopic equation, Eq.~(\ref{eqn:red_eqR}) together with Eq.~(\ref{eqn:fsolution}), yielding
\begin{equation}
\label{eqn:ms_R4}
 -\partial_\tau \langle x_\alpha \rangle
    = \int \dd  \vec{x} \,   x_\alpha \mathcal{L'} h(\vec{x}).
\end{equation}
We evaluate the corrections associated with the substrate, $\alpha<N$, to leading order in $x$ and $\nu$.
Defining the transfer matrix $\underbar{T}_{\alpha,i}=K_2^i (\delta_{i+1,\alpha}-\delta_{i\alpha})$, one can show the following results:
\begin{align}
  \int \dd \vec{x} \,  \vec{x} & \mathcal{L}_3 h = \underbar{E}\langle\vec{x}\rangle, \\
-\int \dd \vec{x} \,  \vec{x} & \sum_{ij}(\chi \underbar{M}^{-1} \varphi^T)_{ij} h  \nonumber \\
&  = (\underbar{1} +  \underbar{T}\, \underbar{M}^{-1}) \underbar{E}\langle\vec{x}\rangle +\frac{\nu}{\gamma}\underbar{M}\vec{S},\\
\int \dd \vec{x} \,  \vec{x} & \sum_{ij} (\chi \underbar{S} \chi^T)_{ij} h  = \frac{\nu}{\gamma} \left(\underbar{M}+ \underbar{T}\right)\vec{S},
\end{align}
where $\vec{S}$ is the vector obtained by summing over all rows of the matrix $\underbar{S}$, i.e., $(\vec{S})_i=\sum_{j}S_{ij}^T$. Summing up these equations, we obtain
\begin{align}
 \partial_\tau \langle \vec{x}\rangle= \underbar{T}\, \left(\underbar{M}^{-1} \underbar{E}\, \langle\vec{x}\rangle-\frac{\nu}{\gamma}\vec{S}\right),
\end{align}
which has the steady state solution $\langle \vec{x}\rangle=({\nu}/{\gamma})\underbar{E}^{-1}\underbar{M}\vec{S}$.
Note that the result is virtually independent of $\gamma$, if we rewrite
\begin{align}
 \label{eqn:networkshift} \langle x_\alpha \rangle ={\nu}\sum_{i}(\underbar{E}^{-1}\underbar{D}\,\underbar{J}^{-T})_{\alpha i}.
\end{align}
Hence it is clear that the finite volume corrections for substrate species are non-zero; this indeed signals the  breakdown of the law of mass action on mesoscopic length scales or equivalently for low copy number of molecules.

It can also be shown that the elements of the inverse Jacobian of the complex are given by
\begin{align}
\label{eqn:jac_exp}
 J_{ij}^{-1}=\frac{1}{K_M^j}
      \left( \delta_{ij} - \frac{a_i G_{ij}}{1 + \hat{\vec{e}}_i^T \underbar{a}\hat{\vec{e}}_i} \right),
\end{align}
where $\underbar{a}=\text{diag}([A_i]/K_M^i)$ and $a_i$ is the $i^{th}$ diagonal element of the latter matrix with value $[A_i]/K_M^i$ (the reduced macroscopic substrate concentration of species $i$). Given the definition, Eq.~(\ref{eqn:Gdefn}), it is straightforward to verify from the above equation that the elements of the inverse Jacobian are always positive valued. Since both matrices $\underbar{D}$ and $\underbar{E}^{-1}$ are also positive it then follows that the finite volume corrections to the substrate concentration, Eq.~(\ref{eqn:networkshift}), are always positive. In other words, the predictions of the REs will always underestimate the steady-state substrate concentrations for a substrate-enzyme network confined in a small volume.

The equations for the finite volume corrections can be conveniently expressed in a form which is useful for obtaining insight about the physical origin of the corrections and also for their numerical computation. We write the matrix $\underbar{D}=\underbar{A}\,\underbar{G}\,\underbar{C}$, where $\underbar{A}=\text{diag}([A_i])$ and $\underbar{C}=\text{diag}([C_i])$. The Jacobian of the free enzyme is simply the transpose of the Jacobian of the complex such that $(\vec{J}_E^{-1})_i=\sum_j J_{ji}^{-1}$; this can be directly computed using Eq.~(\ref{eqn:jac_exp}). These two equations together with the steady-state condition, Eq.~(\ref{eqn:MFss_cond}), written here as $\underbar{C}=\underbar{E}\,\underbar{a}$, the identity $(\underbar{E}^{-1}\, \underbar{G}\, \underbar{E})_{\alpha j}=(\vec{\hat{e}}_\alpha)_j$ and Eqs.~(\ref{eqn:dens_pred_stoch}) and (\ref{eqn:networkshift}), allow us to write a simple, final equation for the steady-state substrate concentration as predicted by the stochastic model
\begin{align}
 \label{eqn:r_err}
 \frac{\rho_\alpha}{[A_\alpha]}=1+R_{\alpha},
\end{align}
where
\begin{equation}
\label{eqn:defn_R}
R_{\alpha} = \frac{1}{V} \vec{\hat{e}}^T_\alpha \underbar{a} \, \vec{J}_E^{-1}.
\end{equation}
The parameter, $R_{\alpha}$, is proportional to the product of the relaxation time of the enzyme species and the corresponding reduced macroscopic substrate concentration, summed over all substrates which can bind to the same enzyme. It is also straightforward to derive expressions for the relative error made by the RE model, for the absolute differences between the reduced mesoscopic and macroscopic concentrations and for the absolute differences between the sum of the reduced mesoscopic and macroscopic concentrations, respectively
\begin{align}
\label{eqn:Rerror}
&R_\text{error}^{\alpha} =\frac{\rho_{\alpha} -[A_{\alpha}]}{\rho_{\alpha}}= \frac{R_{\alpha}}{1+R_{\alpha}}, \\
\label{eqn:abs_ind}
&\frac{\rho_{\alpha}-[A_\alpha]}{K_M^\alpha} = R_{\alpha} a_{\alpha}, \\
\label{eqn:abs_sum}
&\sum_{\alpha} \frac{\rho_{\alpha}-[A_\alpha]}{K_M^\alpha} = \sum_{\alpha} R_{\alpha} a_{\alpha},
\end{align}

In the following subsections, we will apply the general results developed so far, to three different cases which are commonly encountered in biology. We will verify the main theoretical predictions, i.e., Eqs.~(\ref{eqn:Rerror})-(\ref{eqn:abs_sum}), by comparison with detailed stochastic simulations.

\subsection{Single-substrate reaction with single-enzyme species}
\begin{figure}[h]
 \centering \includegraphics[width=0.42\textwidth]{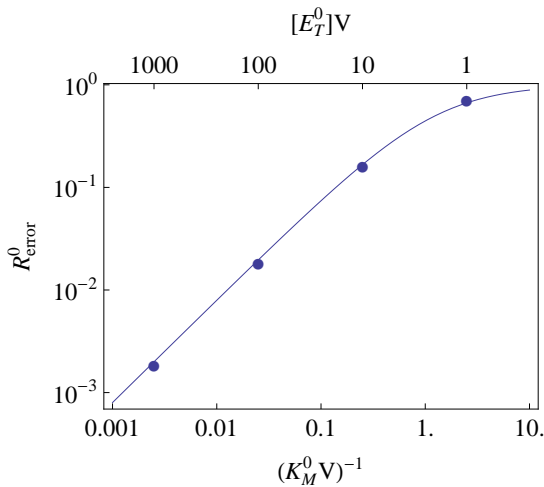}
\caption{\emph{Single-substrate, single-enzyme reaction.} Finite volume scaling of the relative error in the substrate concentration as predicted by the RE model, i.e. $R_\text{error}^0=(\rho_0 - [A_0])/\rho_0$. The error stems from a finite-volume correction to the mean concentration due to intrinsic noise. The solid line shows the theoretical estimate as given by Eq.~(\ref{eqn:Rerror}) together with Eq.~(\ref{eqn:type0_rho}) in the text. The data points are obtained from simulations carried out using Gillespie's stochastic simulation algorithm under steady-state conditions. The parameters are: $K_M^0=1/5$ ($k_1=5$, $k_2^0=0.5$, $k_{-1}^0=0.5$) and $[E_T^0]=0.5$. Timescale separation is guaranteed since $\gamma=([A_0]+K_M^0)=10 [E_0]$ and near-saturation conditions ensue since $k_\text{in}/(k_2^0 [E_T^0])=0.8$ (a value of 1 indicates complete saturation). The upper x-axis indicates the corresponding discrete number of enzyme molecules $[E_T^0]V$ used in the simulation. Note that the agreement between theory and simulation is excellent even when there is just one enzyme molecule in the compartment.}
\label{fig:type0_voldep}
\end{figure}

The reaction where a single enzyme-species catalyzes only a single-substrate is the classical Michaelis-Menten reaction textbook example
\begin{align}
 \label{eqn:example1}
 &\overset{k_{in}}{\longrightarrow} A_0, \quad {E}_0+A_0\xrightleftharpoons[k^0_{-1}]{k_1} C_0 \overset{k^0_2}{\longrightarrow} {E}_0+A_{1}.
\end{align}
The enzyme only binds to a single substrate species and hence $G_{00}=1$. It then follows that Eq.~(\ref{eqn:defn_R}) evaluates to
\begin{align}
\label{eqn:type0_rho}
R_0 = \frac{1}{K_M^0 V}\frac{a_0}{(a_0+1)},
\end{align}
where $a_0=[A_0]/K_M^0$. Substituting the above in Eq.~(\ref{eqn:r_err}) gives an expression for the ratio of the mesoscopic and macroscopic substrate concentrations; this agrees exactly with that obtained by taking the limit of well-separated time scales (i.e., $\gamma=([A_0]+K_M^0) \gg [E_0]$) of the finite-volume correction previously calculated by Grima\cite{Grima2} using the system-size expansion including terms of order $V^{-1/2}$. Note that the finite volume corrections are significant for large steady-state substrate concentrations, i.e., when the enzyme is working in saturated or near-saturated conditions. The predicted dependence of the relative error [i.e., Eq.~(\ref{eqn:Rerror}) together with Eq.~(\ref{eqn:type0_rho})] on the size of the finite volume was tested using stochastic simulations (Fig.~\ref{fig:type0_voldep}). Numerics and theory agree very well over at least $3$ orders of magnitude of the compartment volume, or equivalently over the whole typical physiological range of enzyme copy numbers (from $1$ to $1000$).

\subsection{Multi-substrate network with single-enzyme species}

We now consider a network where different substrates compete for catalysis by a single-enzyme species
\begin{align}
 \label{eqn:example2}
 \notag
 &\overset{k_{in}}{\longrightarrow} A_0, \quad {E}_0+A_0\xrightleftharpoons[k^0_{-1}]{k_1} C_0 \overset{k^0_2}{\longrightarrow} {E}_0+A_{1}, \notag \\ &{E}_0+A_1\xrightleftharpoons[k^1_{-1}]{k_1} C_1 \overset{k^1_2}{\longrightarrow} {E}_0+A_{2}, \notag \\ & ... \notag \\   &{E}_0+A_{N-1}\xrightleftharpoons[k^{N-1}_{-1}]{k_1} C_{N-1} \overset{k^{N-1}_2}{\longrightarrow} {E}_0+A_{N}.
\end{align}

Such type of reactions are commonly associated with the methylations of proteins, DNA and RNA by the methyltransferase class of enzymes. A strictly unidirectional methylation occurs for instance in the nitrogen trimethylation of Lys\cite{Walsh}. A single amino acid, embedded in a protein, can typically be methylated only a few times (e.g., three times in the case of Lys). But the mere large number of amino acids composing proteins means that the number of methylations can generally be very large. The modification of the protein from the unmethylated (or even partially methylated) form $A_0$ to the fully methylated form $A_N$ is processed by the enzyme within $N$ catalytic steps. Partially methylated intermediate states are then described by $A_1, \dots, A_{N-1}$. This type of network has also been proposed as a candidate for explaining the kinetics of protein digestion.\cite{Schnell}

The enzyme can now bind to all the substrate species and hence $G_{ij}=1$. It also follows that since we have one enzyme species then $[E_T^i]=[E_T^0]=[E_T]$. It can then be shown that Eq.~(\ref{eqn:defn_R}) evaluates to
\begin{equation}
\label{eqn:type1_rho}
R_i = \frac{1}{V} \sum_j \frac{1}{K_M^j}\frac{a_j}{a+1},
\end{equation}
where $a = \sum_{i} a_{i}$ is the sum of reduced macroscopic concentrations. To obtain further insight into the above equation and to evaluate Eqs.~(\ref{eqn:Rerror})-(\ref{eqn:abs_sum}) it is necessary to obtain expressions for the reduced macroscopic substrate concentrations. Using the REs, Eq.~(\ref{eqn:meanfield}), and applying the QSSA to the complex concentration variables we find that the reduced macroscopic equations are
\begin{align}
\label{eqn:type1_mf}
\frac{d[A_0]}{dt}=k_\text{in}-v^0, \ \
\frac{d[A_i]}{dt}=v^{i-1}-v^i,
\end{align}
where $v^i=k_2^i[E_T] \ a_i/(1+a)$, i.e., the catalytic velocity of step $i$. Imposing the steady-state condition on the macroscopic substrate concentration equations, Eq.~(\ref{eqn:type1_mf}), we find that the catalytic velocities of all steps must be the same and equal to the input rate, $v^i=k_\text{in}$. It follows that the reduced macroscopic substrate concentrations are $a_i = \eta_i (1-\eta)^{-1}$ and their sum is $a = \eta (1-\eta)^{-1}$ where $\eta_i$ is the dimensionless parameter defined as
\begin{align}
\label{defn_eta}
  \eta_i = \frac{k_\text{in}}{[E_T] k_2^i},
\end{align}
and $\eta = \sum_i \eta_i$. Note that $\eta_i$ is the ratio of the input rate and of the maximum rate at which the enzyme can catalyze substrate $A_i$ into substrate $A_{i+1}$. Similarly $\eta$ is the ratio of the input rate and of the overall maximum rate at which the enzyme can catalyze the initial substrate $A_0$ into the product $A_N$. Hence both $\eta_i$ and $\eta$ have a value between $0$ and $1$. It follows also that $\eta$ is a measure of enzyme saturation.

Given the above results one can now evaluate Eqs.~(\ref{eqn:abs_ind}), (\ref{eqn:abs_sum}), and (\ref{eqn:type1_rho}) which leads to
\begin{align}
\label{eqn:type1_rho_a}
&R_i=\frac{1}{V} \sum_j \frac{\eta_j}{K_M^j}, \\
\label{eqn:type1_abs_ind}
&\frac{\rho_i-[A_i]}{K_M^i} =  R_i \frac{\eta_i}{1-\eta},\\
\label{eqn:type1_abs_sum}
&\sum_i\frac{\rho_i - [A_i]}{K_M^i} = R_i \frac{\eta}{1-\eta}.
\end{align}
From Eqs.~(\ref{eqn:r_err}), (\ref{defn_eta}), and Eq.~(\ref{eqn:type1_rho_a}) one can deduce that: (i) the concentration of each different substrate species is amplified (from the deterministic concentration) by the same factor; (ii) the deviation from the predictions of the deterministic model increases with decreasing volume, decreasing Michaelis-Menten constants, increasing enzyme saturation and increasing network size $N$. The latter stems from the fact that increasing the number of substrates will increase competition for the single enzyme species meaning that the copy number of free enzyme molecules at any one time becomes smaller and hence leads to increased noise-induced effects.

\begin{figure}
 \centering \includegraphics[width=0.49\textwidth]{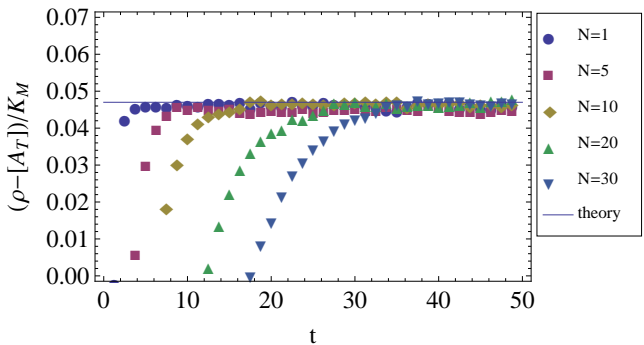}
\caption{\emph{Scaling of the finite-volume corrections with network size for a multi-substrate network with single-enzyme species}. Plot of the absolute difference between the sum of the reduced mesoscopic and macroscopic substrate concentrations versus the network size, $N$, for a homogeneous set of rate constants, $k_{-1}^i = k_{-1}^0$, $k_{2}^i=k_2^0$. Theory predicts that if the input rate is rescaled by the network size, $k_\text{in}=k_\text{in}^0/N$, then the absolute error in steady-state conditions [as given by Eq.~(\ref{eqn:type1_abs_sum})] should be independent of network size. Data points are obtained from stochastic simulations using Gillespie's algorithm (with a rescaling of $k_{\text{in}}$ as mentioned above) and solid lines are the theoretical estimates using Eq.~(\ref{eqn:type1_abs_sum}). Note that in the y-axis label we have used $\rho$ to mean $\sum_i \rho_i$ and $[A_T]$ to mean $\sum_i [A_i]$. Parameter values are found in the text.}
\label{fig:cascade_numerics}
\end{figure}

We have carried out extensive stochastic simulations using the Gillespie algorithm\cite{Gillespie1} to test the accuracy of our predictions. The four tests are as follows:

\emph{Test 1: Dependence of the finite-volume corrections with network size.} We first impose homogeneous rate constants, i.e., a fixed set of rate constants for all enzyme reactions in the network. It then follows that if we make the scaling $k_\text{in}=k_\text{in}^0/N$, the quantity $\eta = \sum_i \eta_i$ will be independent of the network size. Consequently under this scaling, $R_i$, is predicted to be independent of the size and so is the sum of the absolute reduced concentrations, Eq.~(\ref{eqn:type1_abs_sum}). Stochastic simulations show perfect agreement with this prediction (Fig.~\ref{fig:cascade_numerics}) which implicitly proves that the theory correctly predicts that finite volume corrections increase with network size. The relevant simulation parameters are: $V=500$, $k_1=1000$, $k_{-1}^i=0.5$, $k_2^i=20$, $(K_M V)^{-1}=0.1$, $[E_T]V=5$, $k_\text{in}^0=0.1$, $\eta=0.5$ and $\gamma = 10 \ \text{tr}(\underbar{E})$.

\begin{figure}[!t]
\centering \includegraphics[width=0.43\textwidth]{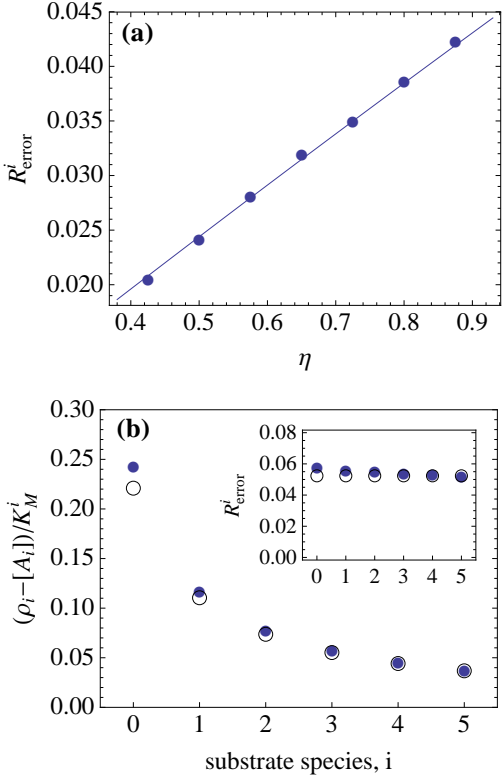}
\caption{\emph{Dependence of the deviations from the RE predictions with enzyme saturation and position in the network for a multi-substrate network with single-enzyme species.} In (a) we plot the relative error as a function of $\eta$ which is a measure of enzyme saturation. In (b) we plot the absolute differences in the reduced mesoscopic and macroscopic substrate concentrations as a function of the substrate species index and hence as a function of position in the six substrate network. The inset shows the relative error for all species. The data points are from stochastic simulation and the solid lines [in (a)] and the open circles [in (b)] are the theoretical predictions. The rate constants are heterogeneous in both cases but the Michaelis-Menten constants for each reaction in the network are the same for (a) while they vary according to position in the network in (b). See the text for parameter values and for a detailed discussion.}
\label{fig:type1_paramdep}
\end{figure}

\emph{Test 2: Dependence of the relative error with enzyme saturation.} We fix the Michaelis-Menten constant, $K_M^i=K_M$, and the size of the network. It then follows from Eqs.~(\ref{eqn:Rerror}) and (\ref{eqn:type1_rho_a}) that the relative error is simply given by $R_{error}^i=\eta/(\eta + K_M V)$ i.e. the relative error increases with increasing enzyme saturation. The prediction is verified by stochastic simulation [Fig.~\ref{fig:type1_paramdep}(a)], where the rate constants were chosen so that $k_2^i = k_2^0$ for $i$ even and $k_2^i = f^{-1} k_2^0$ for $i$ odd and the specific parameter values used were: $V=500$, $k_1=1000$, $N=6$, $f=0.5$, $k_{\text{in}}=1/6$, $k_2^0=10$, $\gamma\geq 10 \ \text{tr}(\underbar{E})$, $(K_M V)^{-1}=0.05$, $[E_T]V=5$.

\emph{Test 3: Dependence of the relative error with species type.} As previously mentioned, an inspection of Eq.~(\ref{eqn:type1_rho_a}), shows that the quantity $R_i$ and hence the relative error $R_{error}^i$ are predicted to be the same for all species in the network. This agrees with the results of stochastic simulation [inset of Fig.~\ref{fig:type1_paramdep}(b)] of a six substrate network where $K_M^i=(1+i)K_M^0$, $k_2^i=(1+i)k_2^0$, $k_{-1}^i=(1+i)k_{-1}^0$, and the specific parameter values were: $V=1000$, $k_{\text{in}}=1/60$, $k_2^0=9$, $k_{-1}^0=1$, $k_1=1000$, $[E_T] V=5$.

\emph{Test 4: Dependence of the finite-volume correction on the position in the network.} Here we test the ability of the theory to predict the absolute difference between the reduced mesoscopic and macroscopic substrate concentrations as predicted by the stochastic and deterministic models, i.e. Eq.~(\ref{eqn:type1_abs_ind}), for each individual species in the network. For rate constants chosen as in Test 3, the differences are predicted to be proportional to $(1+i)^{-1}$, which is confirmed by simulations [Fig.~\ref{fig:type1_paramdep}(b)]. All parameter values are exactly as in the previous test.

\subsection{Multi-substrate network with multi-enzyme species}
Finally we consider a sequential reaction network which is typically associated with metabolism\cite{Tschudy}
\begin{align}
 \label{eqn:example3}
 \notag
 &\overset{k_{in}}{\longrightarrow} A_0, \quad {E}_0+A_0\xrightleftharpoons[k^0_{-1}]{k_1} C_0 \overset{k^0_2}{\longrightarrow} {E}_0+A_{1}, \notag \\ &{E}_1+A_1\xrightleftharpoons[k^1_{-1}]{k_1} C_1 \overset{k^1_2}{\longrightarrow} {E}_1+A_{2}, \notag \\ & ... \notag \\ &{E}_{N-1}+A_{N-1}\xrightleftharpoons[k^{N-1}_{-1}]{k_1} C_{N-1} \overset{k^{N-1}_2}{\longrightarrow} {E}_{N-1}+A_{N}.
\end{align}

This network is characterized by a different enzyme species for each separate catalytic step and thus $G_{ij}=\delta_{ij}$. The finite-volume corrections can be easily derived from Eq.~(\ref{eqn:defn_R}), since the Jacobian of the complex species is instantly diagonal and hence the result is
\begin{equation}
\label{eqn:type2_rho}
R_i = \frac{1}{K_M^i V} \frac{a_i}{(a_i+1)}.
\end{equation}
It is useful to write the above expression in terms of the constants defining the network by using the REs (with the QSSA applied to the complex concentration variables)
\begin{align}
\frac{d[A_0]}{dt}=k_\text{in}-v_M^0, \ \
\frac{d[A_i]}{dt}=v_M^{i-1}-v_M^i,
\end{align}
where $v_M^i=k_2^i[E_T^i] \ a_i/(1+a_i)$ is the catalytic velocity in this case. The connectivity of the network requires that all catalytic velocities must be the same and equal to the input rate, $v_M^i=k_\text{in}$, from which it follows that the reduced macroscopic concentrations are
\begin{equation}
\label{eqn:one_of_last_eqs}
a_i = \frac{\eta_i}{1-\eta_i},
\end{equation}
where $\eta_i = k_\text{in}/ k_2^i [E_T^i]$, a measure of saturation as before. Hence it follows that Eq.~(\ref{eqn:type2_rho}) reduces to the simple form
\begin{equation}
\label{eqn:type2_rho_explicit}
R_i = \frac{1}{K_M^i V} \eta_i.
\end{equation}
It can be deduced from the above equation that the deviation from the predictions of the deterministic model increases with decreasing volume, decreasing Michaelis-Menten constants and increasing enzyme saturation as for the previous network. However unlike the previous case, for the metabolic network we find no dependence on the network size $N$ and the factor $R_i$ is species-specific implying that the concentration of each different substrate species is amplified (from the deterministic concentration) by a different factor. Hence the finite-volume corrections in this network are locally determined whereas in the previously studied network they are determined by global quantities. This indeed highlights the influence of network topology on the finite-volume corrections.

We have carried out a set of numerical tests in order to verify the accuracy of the theoretical predictions for the relative error [as given by Eq.~(\ref{eqn:Rerror}) together with Eq.~(\ref{eqn:type2_rho_explicit})] and for the absolute differences between the reduced mesoscopic and macroscopic concentrations [as given by Eq.~(\ref{eqn:abs_ind}) together with Eqs.~(\ref{eqn:one_of_last_eqs}) and (\ref{eqn:type2_rho_explicit})]. The two tests are carried out on a six substrate network and are as follows.

\begin{figure}
\centering \includegraphics[width=0.43\textwidth]{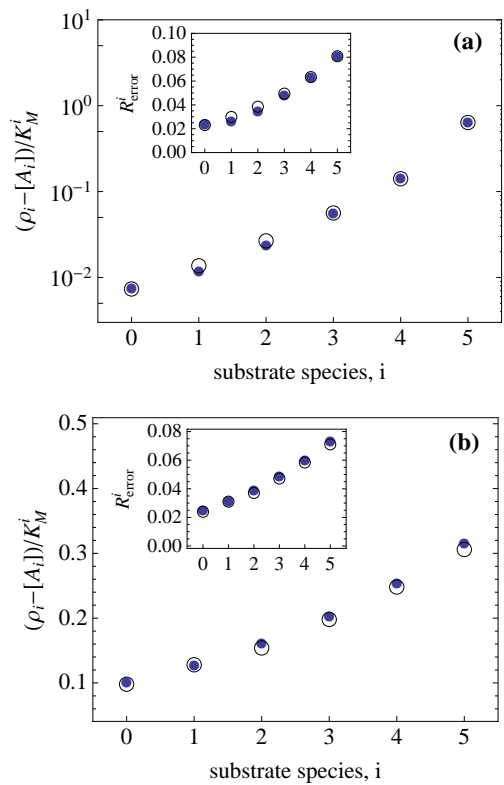}
\caption{\emph{Multi-substrate network with multi-enzyme species.} (a) Heterogeneous catalytic constants and homogeneous Michaelis-Menten constants; (b) Heterogeneous Michaelis-Menten constants and homogeneous catalytic constants. Theoretical predictions are shown as open circles while simulation results are the data points. Theory predicts that the relative error for this network is generally substrate species-specific unlike for the multi-substrate, single-enzyme network where it is the same for all species. The insets confirm this via simulation. For case (b) theory also predicts that the reduced macroscopic concentrations of all substrate species are equal but the reduced mesoscopic concentrations are different for each substrate. This is once again confirmed by simulations which indeed show that the mesoscopic concentrations increase as we go further ``downstream'' in the network. Detailed parameter values can be found in the text.
}
\label{fig:type2_paramdep}
\end{figure}

\emph{Test 1: Heterogeneous catalytic constants and homogeneous Michaelis-Menten constants.} The catalytic constant is chosen to vary with species $i$ by imposing $k_2^i=0.5\times 1.3^{2-i}$ while the Michaelis-Menten constants are all made equal $K_M^i=0.1$. The other relevant parameter values are $k_\text{in}=0.002$, $k_1=10$, $[E_T^i]V=2$, and $V=200$. Figure~\ref{fig:type2_paramdep}(a) shows a comparison between the theory and the stochastic simulations. Note that the relative error for each species is different, in contrast to that for the previous network [inset of Fig.~\ref{fig:type1_paramdep}(b)].

\emph{Test 2: Heterogeneous Michaelis-Menten constants and homogeneous catalytic constants.} The catalytic constants are all made equal, $k_2^i=1$, while the Michaelis-Menten constants are made heterogeneous by choosing the complex dissociation constants to be substrate-specific, $k_{-1}^{i}=9\times 1.3^{2-i}$. The other parameters are $k_\text{in}=0.002$, $k_1=200$, $[E_T^i]V=1$ and $V=400$. Note that in this case the reduced macroscopic concentrations are all equal, $[A_i]/K_M^i=4$, since Eq.~(\ref{eqn:one_of_last_eqs}) only depends on the catalytic constants. In contrast, the reduced mesoscopic concentrations are predicted and confirmed by simulation [Fig.~\ref{fig:type2_paramdep}(b)] to vary from one species to another. This is due to the explicit dependence on the Michaelis-Menten constant in Eq.~(\ref{eqn:type2_rho_explicit}).

\section{Discussion}

In this article we have developed a new approach to the calculation of finite-volume corrections to the substrate concentrations as predicted by the rate equation theory of enzyme-catalyzed networks. This theory is complementary to the general theory of EMREs recently developed by Grima.\cite{Grima3} The EMRE theory gives explicit analytical results only for relatively simple biochemical circuits; we have verified that for one such simple case, i.e., the single-substrate, single-enzyme reaction, the present theory gives the same result as the EMRE. This is an important benchmark for both theories since they were derived using completely different methods (the Poisson Representation and the system-size expansion). A noteworthy achievement of the present theory over the EMRE theory is that it can produce analytical results even for very complex networks involving arbitrarily large numbers of species; this comes about by imposing a separation of timescales of all substrate and complex species in the network which results in a very substantial dimensional reduction of the CME, a result akin to that obtained from applying the QSSA on the rate equations. To our knowledge, this is the first time that the ``mesoscopic QSSA'' has been used to obtain a simple analytical picture of the stochastic dynamics of large networks in finite volumes, i.e., considering effects beyond the conventional fluctuation-dissipation theorem or equivalently beyond those which can be predicted using the linear-noise approximation. Previous studies employing the mesoscopic QSSA have been principally occupied with the need to increase the speed of stochastic simulation.\cite{Arkin, Nemenman2, Peles, Mastny}

In particular, using our approach, we have conclusively shown that the predictions of the rate equation models for the substrate concentrations will always underestimate the real substrate concentrations of a network confined in a finite volume. A generic prediction is that the size of the differences increases with decreasing sub-cellular volume, decreasing Michaelis-Menten constants and increasing enzyme saturation. The topology of the network plays a determining role in the magnitude of these corrections and how they vary according to the ``position'' of the substrate species in the network. For example for the multi-substrate, single-enzyme network we find that the mesoscopic concentrations of all substrate species are larger than the RE predictions by the same factor (a global effect) whereas in the multi-substrate, multi-enzyme network the factor is species specific (a local effect).

We note that in this study we have not considered the effects of diffusion and of finite particle size on the stochastic kinetics; for the case of a simple single-substrate, single-enzyme reaction, these factors have been shown to induce a renormalization of rate constants which is apparent even at deterministic length scales.\cite{Ross} This suggests that the finite-volume corrections that we estimated in this paper constitute only a lower-bound on the actual differences between the real mesoscopic description (which takes into account intrinsic noise and noise stemming from both factors) and the conventional RE description (which ignores both factors and also intrinsic noise).

Concluding, we have presented a complete method of calculation by which one can study the properties of large scale stochastic enzyme kinetic networks. We believe that the ability of the method to give simple expressions describing the kinetics of such large-scale networks is unprecedented and hence may lead to new insight into the effects of noise in biologically relevant networks.

\section*{Acknowledgments} The authors gratefully acknowledge support by the German Science Foundation (P.T. and A.V.S.,  DFG project No.~STR~1021/1-2) and by the Scottish Universities Life Sciences Alliance (R.G.).

\appendix

\section{Derivation of the Poisson representation from the moment generating function equation}
\label{app:rp}
Consider a birth-death Markov process with moment generating function $G(\vec{z})$, which evolves as $\partial_t G(\vec{z})=\mathcal{R}^G G$,
where $\mathcal{R}^G\equiv\mathcal{R}^G(\vec{z},\nabla_z)$. Note that the generic form of $\mathcal{R}^G$ has all the derivatives acting to the right. A complex Poisson representation can be obtained by expanding $G$
\begin{equation}
 \label{eqn:cf}
 G(\vec{z},t)= \int d \vec{\alpha} \, f(\vec\alpha,t) e^{\sum (z_i-1)\alpha_i V}.
\end{equation}
Note that here the summation in the exponent extends over all components of $\vec{z}$ and $\vec\alpha$. The representation is carried out in intensive variables $\alpha_i$. Under the condition that $f$ is sufficiently compact, $f$ satisfies a Liouville equation $\partial_t f(\vec{\alpha})=\mathcal{R} f$, where $\mathcal{R}\equiv\mathcal{R}(\vec{\alpha},\nabla_\alpha)$. The relation between $\mathcal{R}$ and $\mathcal{R}^G$ can be clarified by differentiating (\ref{eqn:cf}) with respect to $t$
\begin{align}
\partial_t G(\vec{z})= \mathcal{R}^G G(\vec{z})
 = \int d \vec{\alpha} \, f(\vec\alpha)
     \mathcal{R}^G(\vec{z},\nabla_z) e^{\sum (z_i-1)\alpha_i V} \\
     = \int d \vec{\alpha} \, f(\vec\alpha)  \mathcal{R}^G(1+V^{-1}\nabla_\alpha,V \vec{\alpha}) e^{\sum (z_i-1)\alpha_i V}.
\end{align}
If $f$ is sufficiently compact, such that by partial integration all boundary terms vanish, we obtain
\begin{align}
\partial_t G(\vec{z}) &= \int d \vec{\alpha}  \,  (\mathcal{R}^G(1-V^{-1}\nabla_\alpha,V\vec{\alpha}) f(\vec\alpha) ) e^{\sum (z_i-1)\alpha_i V}.
\end{align}
Therefore the relation between $\mathcal{R}^G$ and $\mathcal{R}$ is given by
\begin{align}
 \mathcal{R}(\vec\alpha,\nabla_\alpha)\equiv\mathcal{R}^G(1-V^{-1}\nabla_\alpha,V\vec{\alpha}).
\end{align}
Thus formally the Liouville operator $\mathcal{R}$ can be obtained by replacing each $z_i$ by $(1-V^{-1}\partial_{\alpha}^i)$ and each $\partial_z^i$ by $V \alpha_i$ in $\mathcal{R}^G$.

\section{Explicit form of the correlators}
\label{app:transport_coeffs}
The reduced description of enzymatic networks involves the matrix of transport coefficient, defined by $ \underbar{S} \equiv -\langle \vec{z}\mathcal{L}_1^{-1} \vec{z}^T\rangle_\pi$, where the steady-state distribution $\pi$ is obtained from
\begin{align}
 \label{eqn:app_ssc}
 \mathcal{L}_1\pi=0.
\end{align}
The quantity is related to the spectrum of the fast mesoscopic variables $\vec{z}$ described by $\mathcal{L}_1$ at zero frequency. It enters the reduced evolution as a transport coefficient as a consequence of the fast relaxation of the variable $\vec{z}$. To see this we follow the derivation in \cite{Gardiner2} and consider
$ \langle \vec{z} \mathcal{L}_1^{-1} \vec{z}^T\rangle_\pi = \int d\vec{z} \vec{z} \mathcal{L}_1^{-1} (1-P) \vec{z}^T \pi$,
since $P \vec{z}\pi=0$, as required.
Now consider:
\begin{align}
 \int_0^\infty ds \exp (-\mathcal{L}_1 s) = \frac{1}{\mathcal{L}_1} \left. \exp (-\mathcal{L}_1 s)\right|_\infty^0= \mathcal{L}_1^{-1}(1-P),
\end{align}
where we have used that $P=\lim_{s\to \infty}  \exp (-\mathcal{L}_1 s)$. Noticing further that $\exp(-\mathcal{L}_1 s)\vec{z}^T \pi$ is a solution to $-\partial_s f = \mathcal{L}_1 f$ with initial condition $\vec{z}^T \pi$, we obtain:
\begin{align}
 \langle \vec{z} \mathcal{L}_1^{-1} \vec{z}^T\rangle_\pi  =  \int_0^\infty ds \langle \vec{z}(s) \vec{z}^T(0) \rangle_\pi.
\end{align}
Thus if the adiabatic elimination is carried out in the Poisson representation, the transport coefficients are connected only to the spectrum of the mesoscopic Liouvillian $\mathcal{L}_1$, which is independent of the macroscopic shot noise contribution.
Adiabatic fluctuations are necessarily of Ornstein-Uhlenbeck form, as shown in the main text. Therefore we consider (\ref{eqn:app_ssc}) in the form of the potential condition:
\begin{align}
 \left( \underbar{M} \vec{z} - \underbar{D}\nabla_z\right) \pi =0,
\end{align}
from which we can easily verify that $\langle \vec{z}\vec{z}^T\rangle_\pi=-( \underbar{D}\,\underbar{M}^{-1})^T$; the latter is supposed to yield a symmetric expression. Then using the regression theorem, we have:
\begin{align}
 \langle \vec{z} \mathcal{L}_1^{-1} \vec{z}^T\rangle_\pi  &=  \int_0^\infty \dd s\, \exp({-\underbar{M} s})\langle \vec{z} \vec{z}^T \rangle_\pi \nonumber \\
 &= -\underbar{M}^{-1}\underbar{D}^{T}\underbar{M}^{-T}.
\end{align}
Hence it follows that $\underbar{S}=\underbar{M}^{-1}\underbar{D}^{T}\underbar{M}^{-T}$. In a similar manner, one can also prove that $\langle \vec{z}\mathcal{L}_1^{-1} \nabla_{z}^T \rangle_\pi = -\underbar{S}(\underbar{M}^{-1}\underbar{D})^{-T}=-\underbar{M}^{-1}.$

%\newpage

\end{document}